\documentclass[11pt,twoside]{./atmp}

\usepackage{amsmath,amssymb}

\usepackage[all]{xy}
\usepackage{color}
\usepackage{revsymb}
\usepackage{graphics}
\usepackage{graphicx}
\usepackage[tight,footnotesize]{subfigure}
\usepackage{array}

\newbox\tempbox
\newenvironment{nomenclature}{
   \newcommand\entry[2]{
    \setbox\tempbox\hbox{\hspace{0.75in}}
      \hangindent\wd\tempbox\noindent{\makebox[0.75in][l]{##1}}\ignorespaces##2\par}
       }{\par\addvspace{12pt}}

\begin{document}

\title[EHD Based Pulsatile Drug Delivery System]
{Calculation of Droplet Size and Formation Time in Electrohydrodynamic Based Pulsatile Drug Delivery System}


\author[Yi Zheng, et al.]{Yi Zheng*, Yuan Zhang*, Junqiang Hu and Qiao Lin}

\address{Department of Mechanical Engineering, Columbia University, New York, NY 10027}  
\addressemail{Email: yz2308@columbia.edu \\ 
* These authors contributed equally to this work.} 

\begin{abstract}
Electrohydrodynamic (EHD) generation, a commonly used method in BioMEMS, plays a significant role in the pulsed-release drug delivery system for a decade. In this paper, an EHD based drug delivery system is well designed, which can be used to generate a single drug droplet as small as $2.83 nL$ in $8.5 ms$ with a total device of $2\times2\times3mm^3$, and an external supplied voltage of $1500 V$. Theoretically, we derive the expressions for the size and the formation time of a droplet generated by EHD method, while taking into account the drug supply rate, properties of liquid, gap between electrodes, nozzle size, and charged droplet neutralization. This work proves a repeatable, stable and controllable droplet generation and delivery system based on EHD method.
\end{abstract}

\maketitle

\cutpage 


\section*{Nomenclature}\label{S:Nomenclature}
\begin{nomenclature}
\entry{$a,[m]$}{Radius of nozzle}
\entry{$d,[m]$}{Gap between tip of nozzle and ground electrode}
\entry{$E, [V/m]$}{Electric field strength}
\entry{$F_e, [N]$}{Electric force}
\entry{$F_s,[N]$}{Surface force}
\entry{$h,[m]$}{Height of liquid droplet cap}
\entry{$\dot m,[g/s]$}{Drug supply rate}
\entry{$r,[m]$}{Radius of droplet pending at tip of nozzle}
\entry{$r_{eff},[m]$}{Effective radius of droplet}
\entry{$U,[V]$}{Applied external voltage}
\entry{$u,[m/s]$}{Velocity of drug fluid in nozzle}
\entry{$\gamma,[N/m]$}{Surface tension}
\entry{$\varepsilon,[F/m]$}{Drug fluid permittivity}
\entry{$\theta_c,[rad]$}{Contact angle}
\entry{$\mu,[Pa\cdot s]$}{Drug fluid viscosity}
\entry{$\xi,[V]$}{Surface electric potential}
\entry{$\rho_e,[C/m^3]$}{Spatial charge density}
\entry{$\rho_d,[kg/m^3]$}{Drug density}
\end{nomenclature}

\section{Introduction}

Microfluidics and related nano/micro-technology have been active research fields for several years. Numerous microflow devices such as microchannels \cite{1,2}, flow sensors \cite{1,16,17}, microvalves \cite{3}, micropumps \cite{2,5},\cite{9}-\cite{12}, electrowetting \cite{4,6}, electro-hydrodynamics (EHD) \cite{13}-\cite{15},\cite{21}-\cite{24} and magnetic-hydrodynamic \cite{7} have been reported. A droplet dispenser using the microfluidic chip format \cite{8} has recently found increasing popularity due to its potential for massively parallel array ability and scalability of its size \cite{18}. Especially, the latter characteristic reduces the dead volume of the device, which is essential when dealing with an extremely small sample volume.

The pulsed and discontinuous drug delivery systems are more advantageous in the academic and clinic applications, which can be used in therapeutics as concentrations of solutions vary with time, or matches body’s release of peptides and hormones \cite{1,19}. The device would deliver drugs according to various stimuli such as chemical pH, electric fields, and temperature. By judging the degree of external signal, and the device would release appropriate amounts of drug \cite{20,21}.

In this paper, the EHD based drug delivery system is applied mainly for three reasons. Firstly, it doesn’t require complex fabrication because there is no moving part in the whole device. Secondly, it can produce mono-disperse droplets in a wide range of sizes using either electrospray or single-droplet generation. Thirdly, instability, poor directionality, and diversity in droplet size can be reduced by coating Teflon on surface of PDMS chip and reducing surface tension of liquid in fabrication process \cite{21,23}. Our work shows that a stable, repeatable and controllable generation of volume at a constant rate has been achieved by the EHD method. Many types of solutions can be delivered through the device such as DI water, acetic acid, methanol, and so on.

\section{Device Concept and Operating Principle}

\begin{figure}
\includegraphics[width =6.0cm]{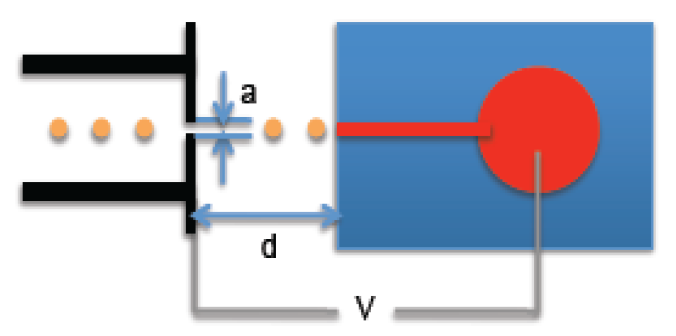}
\caption{\label{fig:1} Schematic of EHD drug delivery device
}
\end{figure}

A typical EHD based micro drug delivery device is shown in Fig. \ref{fig:1}. An electric field would be formed after when we apply voltage on two electrodes. Drug solutions would be under the electric pressure as soon as they arrive at the tip of the red nozzle. Under the pressure, the liquid is elongated and breaks up into single micro-droplets from the tip. Taylor cone, which is very common in hydrodynamic spray processes, is suppressed by controlling the surface wetting property of the PDMS device and the surface tension of the sample liquids.

\section{Model Design and Fabrication}
\subsection{Device Design and Dimensions}
\begin{figure}
\includegraphics[width =10.0cm]{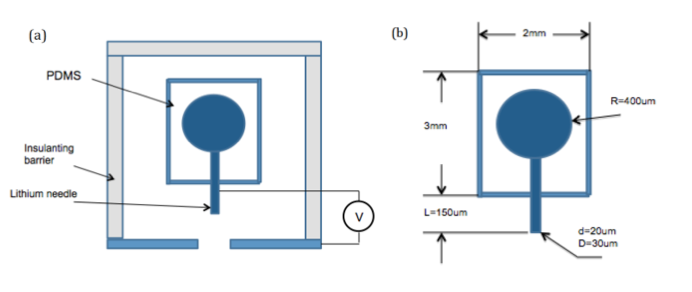}
\caption{\label{fig:2} Design and dimensions of EHD based device. (a) PDMS core (blue), external box (grey) and applied voltage (black); (b) dimensions of PDMS core and metal tube.
}
\end{figure}

The core of the device made of PDMS consists of a spherical drug reservoir and a cylindrical tube extruded from the reservoir, as shown in Fig. \ref{fig:2}. A metal tube/needle, as an anode, is inserted into the PDMS core.
\subsection{Materials}
Electrode: Silicon-based electrode coated with polyimide over an adhesion promoter.\\
Insulating material: Diamond-like carbon (DLC) film adheres strongly to some plastics providing DLC is biocompatible.
\subsection{Fabrication}
The big chunk of PDMS in the middle is supposed to be made up by two half parts. For the reservoir, it is a half sphere, and for the micro-channel, it is a half cylinder. Overall, the fabrication will be complicated. So we choose to use Grey Scale method on SU-8 of E-beam lithography.
\begin{figure}
\includegraphics[width =8.0cm]{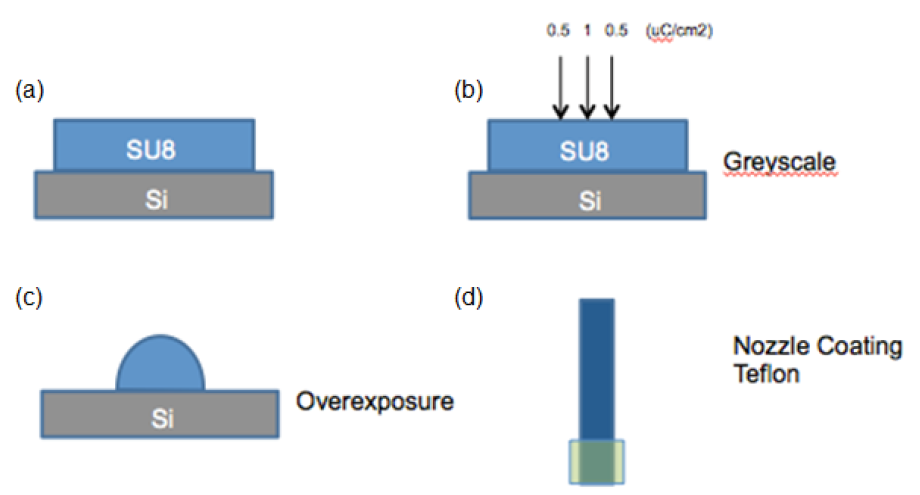}
\caption{\label{fig:3} Fabrication process of drug reservoir and Teflon coating at tip of nozzle
}
\end{figure}
Teflon coated on the inner and outer surface of tip of the metal nozzle, as shown in Fig. \ref{fig:3}(d), can improve the stability and directionality of the droplet formation, and further reduce the droplet size by reducing surface tension of liquid.

Fabrication process, as shown in Fig. \ref{fig:3}, consists of four steps as follows.
(1) Put SU-8 E-beam resist on top of silicon wafer.
(2) Grey Scale to overexpose the middle, and by manipulating the gradient density to get a perfect half cylinder and a half sphere. In Fig. \ref{fig:3}(b), 1 and 0.5 $\mu C/cm^2$ , is just a trial number.
(3) After overexposure, one half core is done. Then cast PDMS on top of it.
(4) Coat metal nozzle with Teflon.

\section{Charged Droplet Neutralization}
\begin{figure}
\includegraphics[width =8.0cm]{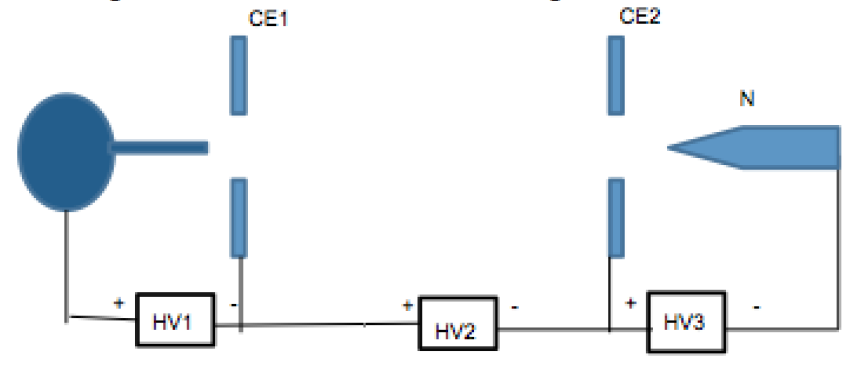}
\caption{\label{fig:4} Schematic of a neutralization system
}
\end{figure}

Droplets generated by EHD method would carry a slight amount of charges. Since nobody has really taken the droplet neutralization effect into account for a matured drug delivery system,  this part is hard to accomplish. Here, we give out three possible solutions for this neutralization problem. A neutralization system is shown in Fig. \ref{fig:4}.
\subsection{Negative Corona Discharge}
When the charged droplet comes out of CE1, by changing HV3, we can actually change the negative electrons released from tip N, which are used to neutralize the positive charges carried on droplet \cite{25}.
\subsection{Negative Spraying of Volatile Liquid}
By replacing the N part with a capillary of spraying volatile liquid out, which will bring negative electrons in the liquid, when the evaporation is finished, electrons remain to neutralize the positive charges.
\subsection{Electron Cloud}
Electrons generated at the wall by proton losses are accelerated and decelerated by the beam potential and hit the opposite wall with a net energy gain, producing secondary electrons and forming a charged cloud.

\section{Calculation of Droplet Size and Formation Time}
\subsection{Droplet Size}
Droplets are formed at the tip of nozzle, that is, the metal tube inserted into PDMS channel. Figure \ref{fig:5} below shows the formation process of a droplet.
\begin{figure}
\includegraphics[width =9.0cm]{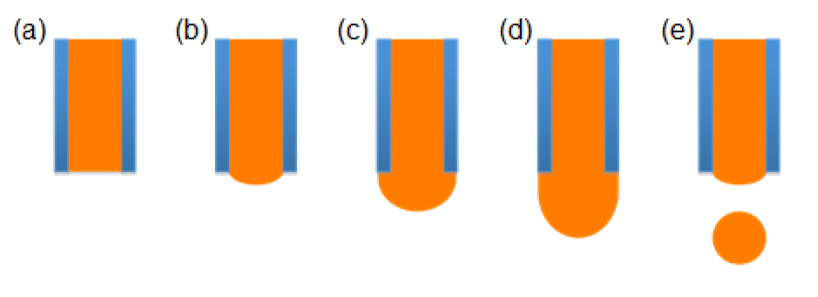}
\caption{\label{fig:5} Formation process of a droplet, (a) ~ (e)
}
\end{figure}
Now, assume the geometry of a liquid drop as a spherical cap. It is well known that the volume of a liquid droplet formed at the tip is given by
\begin{equation}
V_{liquid}=V_{cap}=\frac{\pi}{3}h^2(3r-h^2)
\end{equation}
where, $h$ is height of a droplet in the larger semisphere, $h=r+\sqrt{r^2-a^2}$, and a is radius of nozzle.

\begin{figure}
\includegraphics[width =3.5cm]{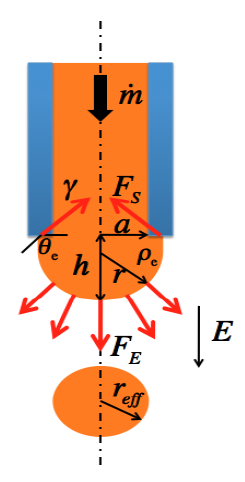}
\caption{\label{fig:6} Parameters used in calculation of a droplet formation
}
\end{figure}

Then, define the spatial charge density, $\rho_e$, in $C/m^3$. The total charge of a liquid drop is
\begin{equation}
Q=\rho_{e}\times V_{liquid}
\end{equation}
In MEMS device, the gravitational effect on a droplet can be neglected. Therefore, there are two groups of forces exerted on a forming droplet, electric force ($F_e$) and surface force ($F_s$), as shown in Fig. \ref{fig:6}.
Electric force caused by the external electric field, due to Coulomb’s law, is given by
\begin{equation}
F_e=E\times Q=E \rho_e V_{liquid}
\end{equation}
where, $E=\frac{U}{d}$, and, $U$ is applied external voltage and $d$ is gap between tip of nozzle and ground eletrode.

Surface force due to the surface tension between liquid, gas and nozzle, is given by
\begin{equation}
F_s=2\pi a \gamma \sin(\pi-\theta_c)
\end{equation}
where, contact angle, $theta_c=\frac{\pi}{2}+\arccos\left(\frac{a}{r}\right)$.

A drug droplet is formed at the tip of nozzle, only when the electric force caused by EHD effect is larger than the surface force on the meniscus of droplet, $F_e>F_s$. The droplet is elongated and breaks up from the nozzle at force equilibrium. The effective radius of a droplet can be calculated as follows.
In equilibrium, $F_e=F_s$, equating Eqs. (3) and (4), volume of a droplet is
\begin{equation}
V_{droplet}=\frac{4}{3}\pi r^3_{eff}=\frac{2\pi a \gamma \sin\theta_c}{E\rho_e},
\end{equation}
and effective radius is
\begin{equation}
r_{eff}=\left(\frac{3 a \gamma \sin\theta_c}{2 E \rho_e} \right)^{1/3}.
\end{equation}

So far, we have calculated the effective radius of a formed droplet in equilibrium at static state. Based on our results, the drop size is dependent on nozzle size, drug solution properties (surface tension), Teflon coating (contact angle), spatial charge density, distance between nozzle and ground electrode, and supplied voltage.

\subsection{Droplet Formation Time}
Two different time constants are of our interests.

$t_{gap}$, the required time between two formed droplets, is given by
\begin{equation}
t_{gap}=\frac{\rho_d V_{droplet}}{\dot m}=\frac{2 \i a \gamma \rho_d \sin\theta_c}{E \rho_e \dot m}
\end{equation}
It significantly depends on the drug supply rate, $\dot m$.

$t_{form}$,the required time to form a single droplet from the tip of nozzle,
\begin{equation}
t_{form}=\frac{V_{droplet}}{\int_{0}^{2\pi} \int_0^{a} v\left(r,\theta\right) r dr d\theta}=\frac{V_{droplet}}{\int_0^{a} v\left( r\right) 2 \pi r dr}
\end{equation}
where, $v\left(r,\theta\right)$ is fluid velocity profile created at the tip of nozzle. We will show how to calculate the velocity profile in the next section.

\subsection{Velocity Profile inside Nozzle}
\begin{figure}
\includegraphics[width =7.0cm]{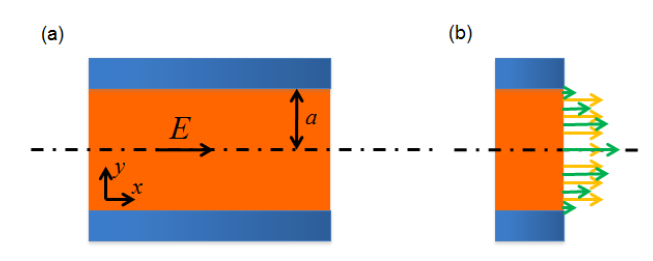}
\caption{\label{fig:7} EOF model. (a) drug flow under constant electric field; (b) two possible velocity profiles at outlet, parabolic (green) and constant (yellow) velocity.
}
\end{figure}

In order to estimate the flow rate inside the electrode induced by an external voltage, we use a well established model called Electro-osmatic flow (EOF), which is the motion of liquid induced by an applied electric field across a charged micro-channel, as shown in Fig. \ref{fig:7}.

Several assumptions are made for the calculation of velocity profile at tip.
(1) Two-dimensional fully developed flow.
(2) Zero pressure gradient.
(3) Slip velocity at channel wall is $u=\frac{\varepsilon \xi}{\mu} E$, where, $\varepsilon$ is drug fluid permittivity, $\mu$ is drug fluid viscosity, and $\xi$ is surface electric potential \cite{26}.

The governing equations used in our calculation are shown below,
\begin{subequations}
\begin{eqnarray}
\triangledown \cdot u&=&0\\
\rho u \cdot \triangledown u&=&-\triangledown p + \mu \triangledown^2u
\end{eqnarray}
\end{subequations}

Write the continuity and momentum equations in Cartesian form, 
\begin{subequations}
\begin{eqnarray}
\frac{\partial u}{\partial x}+\frac{\partial v}{\partial y}&=&0\\
\rho\left(u\frac{\partial u}{\partial x}+v\frac{\partial u}{\partial y}\right)&=&-\frac{p}{\partial x}+\mu\left(\frac{\partial^2 u}{\partial x^2}+\frac{\partial^2 u}{\partial y^2} \right)\\
\rho\left(u\frac{\partial v}{\partial x}+v\frac{\partial v}{\partial y}\right)&=&-\frac{p}{\partial y}+\mu\left(\frac{\partial^2 v}{\partial x^2}+\frac{\partial^2 v}{\partial y^2} \right)
\end{eqnarray}
\end{subequations}

Simplify the equations, we have
\begin{equation}
v=0 \rightarrow \frac{\partial u}{\partial x}=0 \rightarrow u=u(y).
\end{equation}

Using the assumption of zero pressure gradient, Eq. (10c) vanishes and Eq. (10b) can be simplified as:
\begin{equation}
0=\mu\frac{\partial^2 u}{\partial y^2}.
\end{equation}

A general solution for Eq. (13) is in a form of
\begin{equation}
u=C_1y+C_2
\end{equation}
Here, $C_1$ and $C_2$ can be calculated, due to the boundary conditions, from
\begin{subequations}
\begin{eqnarray}
u(-a)=-\frac{\varepsilon \xi}{\mu}E=-C_1 a+C_2\\
u(a)=-\frac{\varepsilon \xi}{\mu}E=C_1 a+C_2
\end{eqnarray}
\end{subequations}
Therefore, $C_1=0$ and $C_2=-\frac{\varepsilon \xi}{\mu}E$.

The fluid velocity is given by
\begin{equation}
u(y)=-\frac{\varepsilon \xi}{\mu}E.
\end{equation}
It is a constant profile, independent of position along y-direction in the nozzle, as the yellow arrows shown in Fig. \ref{fig:7}(b).

It would be much easier to calculate the formation time as velocity is constant, since no integration is needed. Now, substituting Eqs. (5) and (15) into Eq. (8), we can get the formation time of a droplet as
\begin{equation}
t_{form}=\frac{V}{u A}=\frac{V_{droplet}}{\frac{\varepsilon \xi}{\mu}E\pi a^2}=\frac{2 \gamma \mu d^2 \sin\theta_c}{a U^2 \rho_e \varepsilon \xi}.
\end{equation}
Eq. (16) suggests the formation time of a droplet is a function of surface tension, viscosity, gap between electrodes, contact angle, nozzle size, applied voltage, permittivity, etc.

\section{Simulations and Results}
\begin{figure}
\includegraphics[width =12.0cm]{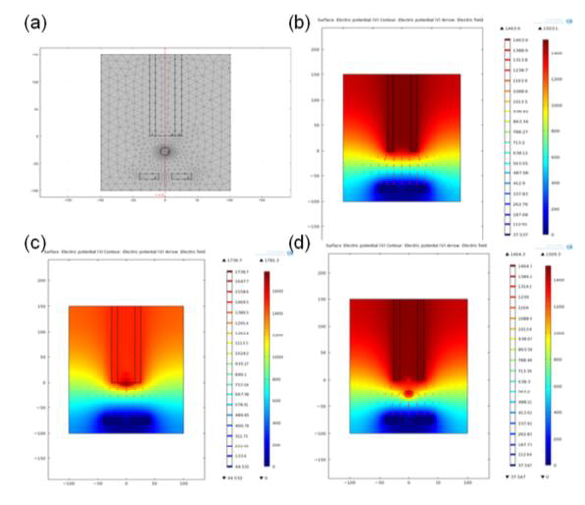}
\caption{\label{fig:8}} COMSOL simulation of electric field distribution between metal needle and ground electrode. (a) numerical analysis domain. (b) initial setup. (c) beginning of droplet formation process. (d) a droplet formed by EHD.
\end{figure}

Here, nozzle size $a=10 \mu m$, applied voltage $U=1500 V$, gap $d=100 \mu m$. So, $E=1.5\times10^7 V/m$. The liquid used in the simulation is water, the common solution to most drug solutions. Its surface tension $\gamma=72\times10^{-3} N/m$, and after Teflon coating, the contact angle increases, $\theta_c=110^o$. Assume the spatial charge density of a droplet . Simulations of an electric field distribution and a droplet formation process in this EHD based system are shown in Fig. \ref{fig:8}.

\begin{figure}
\includegraphics[width =9.0cm]{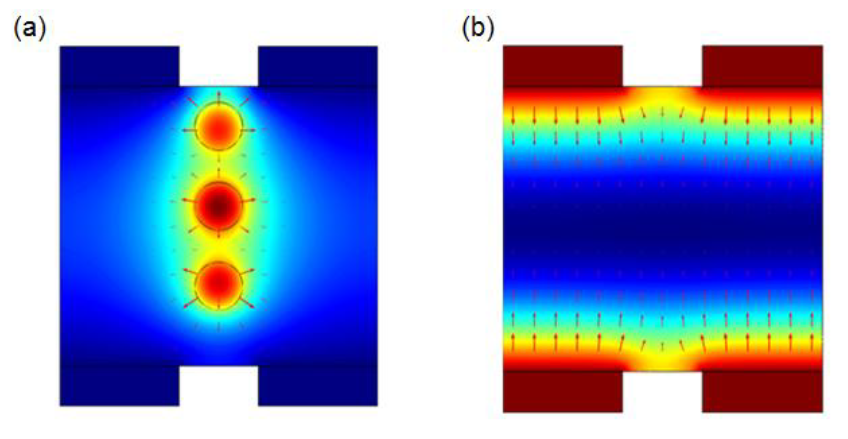}
\caption{\label{fig:9}} Simulation of the neutralization region. (a) high charged droplets without neutralization. (b) neutralization region with opposite charges.
\end{figure}

We also simulated the neutralization process, which is necessary for a newly formed droplet before ejecting into the body. Highly charged droplets formed by EHD method and the neutralization region with opposite charges are shown in Fig. \ref{fig:9}.

\begin{figure}
\includegraphics[width =12.0cm]{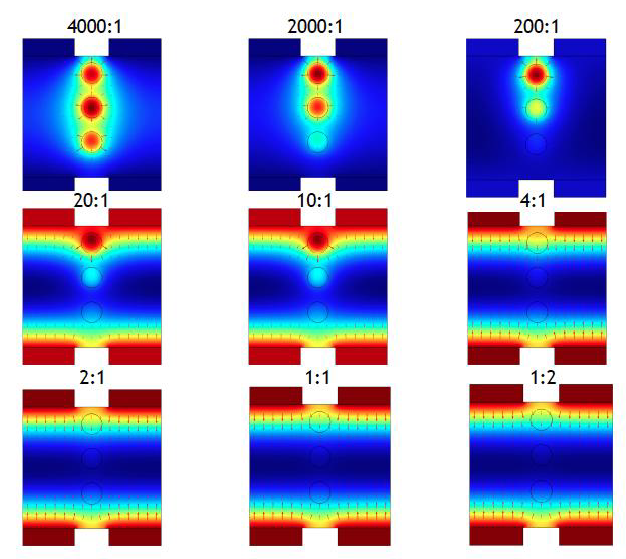}
\caption{\label{fig:10}} Effect of neutralization due to ratios of spatial charge density on droplets and neutralization region
\end{figure}

Different ratios of the spatial charge density on droplets and neutralization region, affect the neutralization effect, as shown in Fig. \ref{fig:10}. The ratio ranges from 4000:1 to 1:2. Neutralization works well when the ratio is around 200:1, at which a formed droplet is neutralized completely. Otherwise, the charges generated by EHD on droplet are either incompletely neutralized or changed to the opposite because of the higher charge density in neutralization region.

\section{Discussion}
This EHD based drug delivery system can deliver a single droplet as small as $2.83 nL$ in $8.5 ms$ with a total device of $2\times 2\times 3 mm^3$, and a working voltage of $1500 V$. The droplet generation is repeatable and controllable.

The dimensions of device and data of the drug delivery system used in calculation are shown in Table 1.
\begin{figure}
\includegraphics[width =8.5cm]{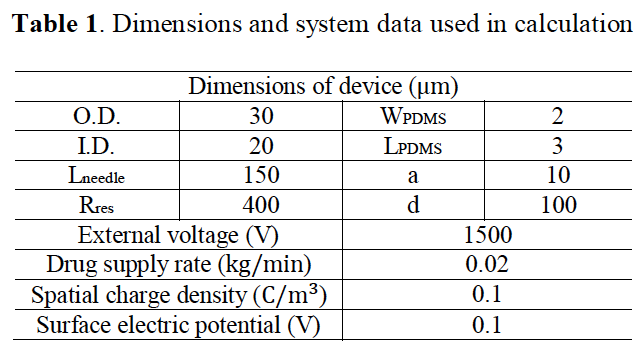}
\end{figure}
Except for DI water used in calculation and simulation, many other solutions could be used for this design. Properties of various solutions can be found in Ref.\cite{21}.

Formation time of a droplet, in Eq. (16), consists of the applied voltage, gap between tip of nozzle and ground electrode, nozzle size, contact angle, and some fluid properties, such as viscosity, permittivity and surface electric potential. For a perticular drug fluid, its properties are fixed. Therefore, the factors to change the formation time are dimensions of the system.

Figure 11 shows the different effects on formation time caused by the four easily manipulated parameters in experiments, external supplied voltage $U$, gap between two electrodes $d$, nozzle size $a$ and contact angle $\theta_c$. In Fig. 11(a) and 11(c), $t_{form}$ reduces as $U$ and $a$ increases separately. It means that it is easier and faster to form a droplet due to EHD at a higher voltage or a smaller nozzle size. In Fig. 11(b), $t_{form}$ increases when $d$ grows, resulting in a decreasing electric field strength $E$. Contact angle increases after Teflon coating. In Fig. 11(d), only when $\theta_c>90^o$, it can reduce $t_{form}$.
\begin{figure}
\includegraphics[width =11cm]{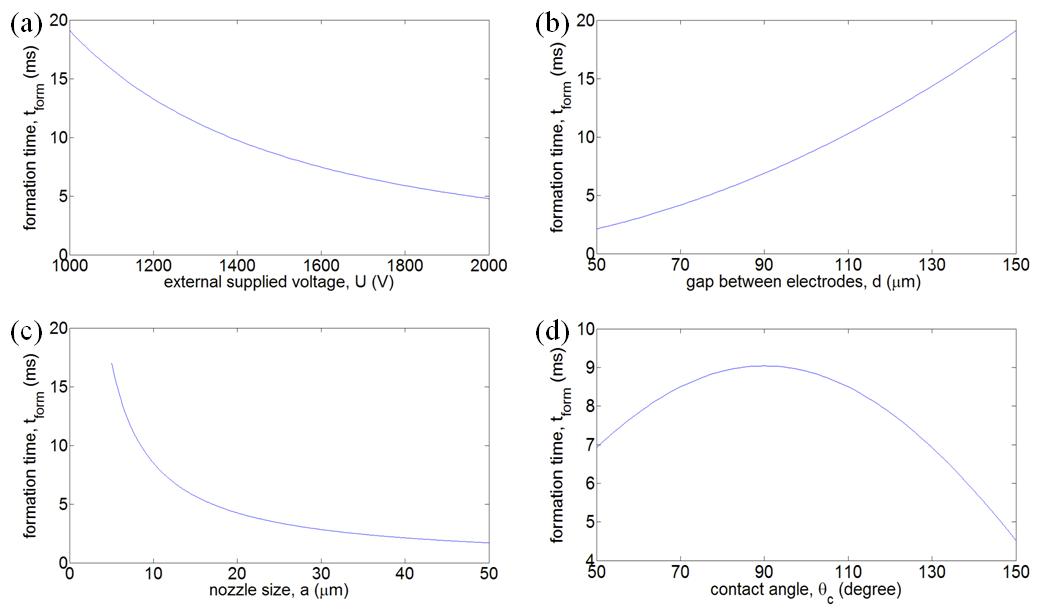}
\caption{\label{fig:11}} Effects on formation time $t_{form}$ caused by external supplied voltage $U$, gap between electrodes $d$, nozzle size $a$ and contact angle $\theta_c$
\end{figure}

\section{Conclusion}
Electrohydrodynamic generation, a commonly used method in BioMEMS, plays a significant role in the pulsed-release drug delivery system for a decade.

In this paper, an EHD based drug delivery system is well designed for the stable, repeatable and controllable generation of a nano/micro droplet at a constant rate. Theoretically, we derive the expressions for the size and the formation time of a droplet generated by EHD method, while taking into account the mass supply rate and properties of drug, gap between two electrodes, nozzle size, and charged droplet neutralization.

Even with an inserted metal tube or needle to enhance the efficiency and stability of the droplet formation, its fabrication and critical alignment remain challenging.





\bibliographystyle{my-h-elsevier}

\end{document}